\PassOptionsToPackage{english}{babel}
\documentclass[useAMS,usenatbib]{mn2e}

\usepackage[english]{babel}
\usepackage{color}
\usepackage{graphicx}
\usepackage{bbold}
\usepackage{subfigure}
\usepackage{breqn}

\title[Circumplanetary disk comparison between core-accretion and disk instability]{Circumplanetary disks around young giant planets: a comparison between core-accretion and disk instability}
\author[Szul\'agyi et al.]{J. Szul\'agyi$^{1}$\thanks{E-mail:
judits@phys.ethz.ch},  L. Mayer$^{2}$ \& T. Quinn$^{3}$\\
$^{1}$ ETH Z\"urich, Institute for Astronomy, Wolfgang-Pauli-Strasse 27, CH-8093, Z\"urich, Switzerland\\
$^{2}$ Center for Theoretical Astrophysics and Cosmology, Institute for Computational Science, University of Z\"urich,\\ 
Winterthurestrasse 190, CH-8057 Z\"urich, Switzerland\\
$^{3}$Astronomy Department, University of Washington, Box 351580, Seattle, WA 98195, USA}
\begin{document}

\date{Accepted XX. Received XX; in original form 2016 June XX}

\pagerange{\pageref{firstpage}--\pageref{lastpage}} \pubyear{2016}

\maketitle

\label{firstpage}

\begin{abstract}
Circumplanetary disks can be found around forming giant planets,
regardless of whether core accretion or gravitational instability built the planet. We carried out state-of-the-art hydrodynamical simulations of the circumplanetary disks for both formation scenarios, using as similar initial conditions as possible to unveil possible intrinsic differences in the circumplanetary disk mass and temperature between the two formation mechanisms.
 
We found that the circumplanetary disks mass linearly scales with the
circumstellar disk mass. Therefore, in an equally massive protoplanetary disk, the circumplanetary disks formed in the disk instability model can be only a factor of eight more massive than their
core-accretion counterparts. On the other hand, the bulk circumplanetary disk
temperature differs by more than an order of magnitude between the two
cases. The subdisks around planets formed by gravitational instability
have a characteristic temperature below 100\,K, while the core
accretion circumplanetary disks are hot, with temperatures even
greater than 1000\,K when embedded in massive, optically thick protoplanetary disks. We explain how this difference can be understood as the natural result of the different formation mechanisms. We argue that the different temperatures should persist up to the point when a full-fledged gas giant forms via disk instability, hence our result provides a convenient criteria for observations to distinguish between the two main formation scenarios by measuring the bulk temperature in the planet vicinity.
% max 250 words % current: 220
\end{abstract}


\section{Introduction}

During the last stage of giant planet formation, a disk forms around
the gas-giant which regulates the gas accretion onto the planet and
from which the satellites form. These disks are called circumplanetary
or subdisks; the latter referring to them being embedded in the circumstellar disk. The two widely accepted planet formation theories, core accretion \citep{Pollack96} and gravitational instability \citep{Boss97} both predict that circumplanetary disks form around giant planets \citep[e.g.][]{QT98,CW02,WC10}.

As of yet, there is no observational evidence of a subdisk; therefore, we have to rely on numerical simulations to examine its properties. The observational efforts have just began, e.g., with the Atacama Large Millimeter Array (\citealt{Pineda16} in prep., \citealt{Perez15}); therefore, making predictions for such observations from hydrodynamical models are crucial. Furthermore, the characteristics of the circumplanetary disks are also very important for satellite formation theory, because the timescales and the formation mechanism itself are still undetermined \citep[e.g.][]{CW02,CW06,ME03a,ME03b}.

In work so far, the masses of subdisks formed via gravitational
instability (GI) or core accretion (CA) have been significantly different. 
The GI smoothed particle hydrodynamic simulations of \citet{SB13}
found very massive subdisks, with 25\% of the planetary mass within the circumplanetary disk (CPD). Similarly, \citet{Galvagni12} and \citet{Galvagni14} recovered 0.5 $\mathrm{M_p}$ subdisks. Limitations of these simulations included low 
resolution and short time evolution. Furthermore, they only followed
the collapse of an isolated clump extracted from a global disk
simulation, and therefore neglected further mass accretion and angular
momentum transport from the circumstellar disk. On the other hand, CA
simulations always resulted in orders of magnitude lighter CPDs
(0.1-1\% of the planetary mass). The radiative, 2D models
of \citet{DA03} found a CPD mass of
$10^{-4} \mathrm{M_{Jup}}$\footnote{G. D'Angelo private
communications} for a Jupiter-mass planet, similar to the isothermal
3D simulations of \citet{Gressel13}. The isothermal 3D simulations
of \citet{Szulagyi14} resulted in a CPD mass of $2\times 10^{-4} \mathrm{M_{Jup}}$ around a Jupiter-mass planet, while the radiative 3D simulations of \citet{Szulagyi15} found $ 1.5\times10^{-3} \mathrm{M_{Jup}}$ for the same massive gas-giant. In conclusion, simulations so far found that the GI formed CPDs more than two orders of magnitude more massive than CA formed subdisks.

Regarding the temperature of the CPD, the CA and the GI simulations predict an order of magnitude difference as well. All non-isothermal core-accretion investigations agree that the peak temperature in the inner subdisk is very high. The temperatures,
of course, depend on the resolution of the simulations and on the
treatment of the planetary; therefore, it is not surprising 
that different investigations measured somewhat different peak temperatures. For a Jupiter-mass planet, \citet{AB09b} argued
for $T = 1600$ K at the planet surface (defined at 0.02
$\mathrm{R_{Hill}}$). They found much a higher value, $T = 4500 K$, with a realistic (i.e. smaller) planetary radius. Of course, the temperature also depends on the viscosity --- through viscous heating --- as was pointed out by \citet{DA03}. Their 2D
radiative simulation gave a maximum of $T  = 1500$ K with their highest viscosity case ($10^{16} \mathrm{cm^2 s^{-1}}$) for a Jupiter-mass
planet. The magneto-hydrodynamic simulations of \citet{Gressel13}
studied somewhat lower mass cores, growing the planet 
from 100 $\mathrm{M_{Earth}}$ to 150 $\mathrm{M_{Earth}}$, but already at these low-mass cores the temperature peaked over
1500-2000 K. Similarly,
in the work of \citet{PN05},
the characteristic temperatures in the CPD
were 1000-2000 K.
The highest resolution CA simulation of \citet{Szulagyi16a} found a maximal temperature of 13000 K when 
the resolution was $\sim110000$ km, i.e. 80\% of a
Jupiter-diameter. This temperature, therefore, refers to a layer below
the planetary surface of a young, puffed up protoplanet. All the above
mentioned non-isothermal CA works agreed that the temperature
profile of the subdisk is very steep: from the maximal temperatures
near the planet surface it quickly declines towards the edge of the
subdisk. On the other hand, the GI studies found significantly lower
temperatures in the planet's vicinity. \citet{SB13} had a peak
temperature of only 40 K, while \citet{Galvagni12} obtained
temperatures in the range 50-100 K in the rotationally supported
envelope of the protoplanetary clump which they identified as the
CPD. The latter work was able to follow the clump collapse because of two orders of magnitude higher resolution (a few Jupiter radii) and showed that the inner core of the clump heats up rapidly to temperatures of higher than 1000 K, at which point dissociation of molecular hydrogen begins at the center. In the meantime, the circumplanetary gas remained cold ($<$100 K). These simulations were among the first ones to show that, in GI, very soon after the collapse a clear dichotomy arises in all physical properties between an inner dense, slowly rotating core and an outer extended circumplanetary envelope or disk.

Another important difference between the CA and GI models is the mass of the circumstellar disk. The GI simulations obviously require very massive protoplanetary disks (usually $\sim 0.1-0.5\, \mathrm{M_{solar}}$) where gravitational instability can occur. In contrast, core accretion simulations use very light  circumstellar disks, close to the Minimum Mass Solar Nebula estimate of $\sim 0.01 \mathrm{M_{solar}}$.

The size of the protoplanet is also among the leading differences between the two formation models. In CA, the
CPD formation is studied assuming that a full-fledged giant planet has
already formed; therefore, approximately a Jupiter radius encloses a
mass on the order of Jupiter mass. On the other hand, in the  disk instability the CPD forms while the clump begins to collapse, when
it has a radius as large as 2-5 AU \citep{SB13,Galvagni12,Galvagni14}. This means that the gravitational potential well
is much deeper in the CA simulations relative to the GI ones. As a consequence, in the former case, the accreting gas can release significantly more energy
into heat compared to the second case. This can be understood due to the fact that the accretional luminosity scales inversely proportional to the accretion radius. However, this accretion radius is 1000 times larger at the onset of clump collapse than in the CA model.

An additional potentially important difference between the
non-isothermal simulations of the two formation scenarios is how the
thermal effects are included. The flux limited diffusion
approximation \citep{Kley89,Commercon11} is used in a number of works on CA formed subdisks \citep[e.g.][]{AB09a,AB09b,Szulagyi16a}. This method includes both radiative cooling and the heating of photons produced by the accretion luminosity. In contrast, most published GI studies  \citet{SB13,Galvagni12,Galvagni14} include a radiative cooling model designed to roughly match the radiative losses in 
flux-limited diffusion simulations but which neglects radiative heating via photons produced by highly compressional flows (e.g. shocks) and the effects of radiation
pressure (see e.g. \citealt{Boley10}, \citealt{RW11}). Some works on disk instability include flux-limited diffusion approximation, but with very low resolution \citep{Mayer07,RW11,Mayer16}. 

The 1-2 orders of magnitude difference in mass and temperature would
predict that observationally, the CA and GI formed subdisks could be
distinguished, even if the observations could only set upper limits on
the CPD mass. However, these differences might come from the fact that
the two sets of simulations are significantly different in the initial
parameters. Motivated by these key differences, for the first time we have
run simulations with very similar initial parameters (i.e. comparably
massive circumstellar disk, semi-major axis, planetary mass,
resolution) to unveil the real differences between GI and CA
subdisks. For the GI case we perform the first global 3D radiative
simulations with enough resolution to clearly separate the CPD and
planetary core, and follow the clump collapse to relatively long
timescales in order to study how the subdisk evolves. The CA
calculations are also state-of-the-art computations, as they are
radiative, 3D global disk simulations with mesh refinement, which
makes them one of the highest resolution studies done so far on
circumplanetary disks. If the GI and CA simulations still give
discrepancies in the CPD temperature and mass despite the similar
initial parameters, they can provide criteria to distinguish subdisks around CA and GI formed planets observationally. 

\section[]{Methods}
\label{sec:numerical}

We use two different numerical methods to study CPDs in core-accretion
and disk instability models because the nature of the two problems is
different. We used a finite-volume code that has excellent shock
capturing capabilities to study core accretion, and a Lagrangian code,
which captures well the global disk dynamics and includes
self-gravity, to study disk instability.

\subsection{Core Accretion Simulations}

We performed grid-based, radiative, three-dimensional, hydrodynamic simulations with the JUPITER code \citep{Szulagyi14,Szulagyi15,Szulagyi16a,Borro06,Benitez15}, developed by F. Masset and J. Szul\'agyi. The code has nested meshes and it is based on a higher order Godunov scheme. The nested mesh technique allows having an entire circumstellar disk while zooming into the planet vicinity with higher resolution.  

The radiative module includes a two-temperature approach for the
Flux-Limited Diffusion
Approximation \citep[e.g.][]{Commercon11}. Therefore, the heating
comes from adiabatic compression and viscous heating, while the
cooling is through adiabatic expansion and radiation
(grey-approximation). We used the gas and dust opacities
of \citet{BL94}; therefore, despite the one-fluid (gas) simulation,
the dust contribution to the temperature is taken into account. The
dust-to-gas ratio was chosen to be 0.001, .i.e., ten times less than
the interstellar medium value. This was motivated by the fact that at
this evolutionary stage of planet formation, most of the circumstellar
disk dust has already been aggregated into larger grains and
planetesimals, lowering the opacity of the
disk \citep[e.g.][]{Ormel09,Ormel11}. The mean molecular weight was
set to 2.3, which is the solar mixture value. The equation-of-state is
ideal gas, $P=(\gamma-1)e$, where the adiabatic index
($\gamma$) is 1.43,  $P$ is the pressure, and $e$ is the
internal energy. We applied a low, constant kinematic viscosity of
$10^{-5} \mathrm{a_{p}}^2\Omega_p$, where $ \mathrm{a_{p}}$ is the
semi-major axis of the planet and $\Omega_p$ is the orbital frequency of the planet. The self-gravity of the gas is not included in these simulations.

Our simulations contain an entire circumstellar disk in the spherical
coordinate system (azimuth, radius, co-latitude) centered on the 
one Solar-mass star. The main parameters of the simulations are in
Table \ref{tab:simus}. The initial surface density profile of the
circumstellar disk was flat, with zero inclination. During the first
150 orbits we run the disk simulation without any planet in it, in
order to reach initial thermal equilibrium of the disk. Then, we
introduce the 10 Jupiter-mass planet through a mass-taper function,
building it up continuously over 50 orbits. After this, the planet
mass is kept constant throughout the simulation. The results are obtained after steady state has been reached (175 orbits after the initial thermal equilibrium and 125 orbits after the planet was fully formed).

\begin{table}
  \caption{Parameters of the core accretion hydrodynamic simulations}
 \label{tab:simus}
  \begin{tabular}{p{3cm}p{1.2cm}p{1cm}p{1.2cm}}
  \hline
     &  CA-1  &   CA-2  &  CA-3\\
 \hline
   CSD-mass [$\mathrm{M_{Sol}}$]	&	0.16	&	0.29	&	0.60 \\
   Planet-mass [$\mathrm{M_{Jup}}$]	&	10	& 	10	& 	10\\
   Semi-major axis  [AU]		&	50	&	5.2	&	5.2 \\
\end{tabular}
\end{table}

The nested meshes were introduced one after the other, until steady state has been reached. In these computationally expensive simulations, we have used four levels of refinement of the base mesh that contains the circumstellar disk. The refined patches include the planet vicinity, and with each level they double the resolution. On the finest level, the cell-diameter was, 0.000576854 code units, which equals to 0.00299964 AU in the simulations where the planet was at 5.2 AU, and 0.0288427 AU when the planet was placed at 50 AU. 

To avoid the singularity of the gravitational potential, we used the
traditional epsilon-smoothing technique, where the gravitational potential is shallwer within an $\epsilon$ distance from the planet:
\begin{eqnarray}
U_p&=&-\frac{G M_p}{\sqrt{x_d^2+y_d^2+z_d^2+{\epsilon}^2}}
\end{eqnarray}
where $x_d=x-x_p$, $y_d=y-y_p$, and $z_d=z-z_p$ are the distances from the planet in Cartesian coordinates, with  $\epsilon$ smoothing length equal to 0.00337907 code units on the finest level, i.e. 6 cell-diameters.  

More details of the JUPITER code, the simulation parameters, and the implementation can be found in \citet{Szulagyi16a} and where similar radiative simulations were carried out on Jupiter-mass planets in low mass circumstellar disks.

\subsection{Gravitational Instability Simulations}

In this paper we also employ a new 3D global disk smoothed particle hydrodynamic (SPH) simulation with unprecedented resolution, which is part of a new simulation suite to be presented in a forthcoming paper (Mayer \& Quinn, in preparation). This is the first simulation that achieves
a resolution of $0.01$ AU in a  200 AU disk, comparable to the resolution of individual clump collapse local simulations \citep{Galvagni12}.
 It employs as many as 42 million SPH particles. We note that the resolution is comparable to that of the core-accretion simulations described in the previous
section for the runs with the planet at large distances (52 AU), which is also the configuration to be compared with the disk instability simulations.

The protoplanetary disk has a mass of $0.6 \mathrm{M_{\odot}}$ and the central star is $1.35 \mathrm{M_{\odot}}$, similar to the host star in the HR8799 system, the prototypical system with 
massive gas giants on wide orbits ($R > 30$ AU) that could have been formed via disk instability.
The disk temperature profile is set up in hydrostatic equilibrium
using a highly accurate iterative procedure that takes into account
full force balance and stellar irradiation at time $t=0$, including
disk self-gravity \citep{Mayer16,RW11}.
The surface density profile is
a power law with an exponent close to $-1$ in the region where fragmentation is expected to happen ($R \sim 30-100$ AU) due to shorter cooling times and low Toomre Q parameter (the minimum Toomre Q drops initially below $1.4$ at R = 60 AU), and has two exponential truncations at the inner and outer 
edge of the disk, which are set at 5 AU and 200 AU, respectively. The central star is treated as a sink particle \citep{RW11}, with a sink radius equal to 4 AU.

The GI simulation presented in this paper was carried out with the new ChaNGa Tree+SPH code, which employs
a CHARM++ parallel programming environment to enable dynamic load
balancing on large supercomputers \citep{Jetley08,Menon15}.  ChaNGa inherits its basic SPH implementation from the GASOLINE and GASOLINE2 codes \citep{Wadsley04,Keller14,Tamburello15}, 
widely used in radiation hydro simulations of 3D self-gravitating
disks \citep[e.g.][]{Mayer02,Mayer07,Mayer16,RW11}. As in GASOLINE2,
ChaNGa employs a modern SPH implementation which uses a geometric
weighting of the density estimate \citep{Keller14,Governato15}
resulting in a formulation of the pressure force analogous to that
presented in \citet{Hopkins13} and \citet{RT01}. Combined with a
turbulent diffusion term in both the momentum and internal energy
equation --- whose formulation is described in \citet{Shen10} --- and
the adoption of an optional Wendland C4 kernel, it avoids artificial
surface tension, resolving the mixing of different fluid phases and physical hydrodynamic instabilities at contact discontinuities.
These new features have been shown to bring SPH in good agreement with finite volume grid-based codes with accurate Riemann solvers \citep{Hopkins14} in modeling the properties of the flow,
while keeping the advantage of a Lagrangian code in modeling disk
dynamics. It provides perfect angular momentum conservation and no
advection errors, which allows the capturing of processes such as ablation of clumps 
by ram pressure, that have never been reported before in either SPH or (fixed) grid simulations (Mayer \& Quinn, in preparation).

Also as in GASOLINE, ChaNGa uses a Monaghan viscosity with $\alpha =
1$ and $\beta = 2$, and a switch to limit the viscosity in purely
rotational flows\citep{Balsara}.
The radiative cooling is based on
local gas properties. For this, we write  the energy loss per time per volume as:
\begin{equation} \label{cool}
\Lambda = (36 \pi)^{1/3} \frac{\sigma}{s} (T^4 - T^4_{\mbox{min}}) \frac{\tau}{\tau^2 + 1}
\end{equation}
where $\tau$ represents the optical depth across a resolution element, $T_{\mbox{min}}$ is the minimum gas background temperature (10\,K), $s=(m/ \rho)^{1/3}$ and $\sigma$ is the Stefan-Boltzmann constant. While equation (\ref{cool}) is only approximate, it allows us to capture the  general behavior of radiative cooling while making the computation much faster than with full-fledged radiative transfer. 
Cooling is most efficient at an optical depth $\tau \sim 1$,
and the two asymptotic limits for large and small $\tau$ recover the dependence of cooling rate on optical depth in the optically thin and 
optically thick limits. This cooling prescription compares reasonably with
flux-limited diffusion calculations, as described in \citet{Boley09} and \citet{Boley10}. As we do not solve the radiation hydrodynamics equation in the diffusion limit, the accretional luminosity of contracting clumps is not included. However, the compressional heating -- generated by PdV work -- and the shock heating is taken into account.

For comparison, and in order to investigate the effect of radiation
physics, we also use another version of this simulation that has 40
times lower mass resolution (gravitational softening  $0.16$ AU) but
includes mono-frequency radiative transfer \citep{Mayer16}. This
simulation of \citet{Mayer16} was carried out with the GASOLINE code
using the implicit method for flux-limited diffusion with photospheric
cooling described in \citet{RW11}, which has been shown to reproduce
expected radiative losses at the disk boundary correctly, a significant improvement over previous methods of disk edge detection in SPH \citep[e.g.][]{Mayer07}.

In order to compute optical depths we used tabulated Rosseland mean and Planck opacities from
\citet{Dal97} and \citet{Dal01} for the gas at solar metallicity
(assuming a dust-to-gas ratio = 0.01).
We included also a variable adiabatic index that takes into account the variation of the ortho/para ratio of molecular hydrogen as a function of temperature, which is important to capture the thermodynamics across spiral shocks in self-gravitating unstable disks \citep{Podolak11}. In order to speed-up further the simulations during the computationally intensive phase of clump collapse we shut-off cooling in the clump core when it has collapsed to about 6 orders of magnitude higher density than the background. This essentially slows down the collapse in the inner region compared to \citet{Galvagni12} but becomes necessary for
computational reasons in order to evolve the disk for longer. We have tested that it has no effect on the circumplanetary disk by running a parallel computation with no cooling shut-off. 
               
\section{Results}

\subsection{The formation of the circumplanetary disk}
\label{sec:cpd_def}      

\begin{figure*}
\includegraphics[width=15cm]{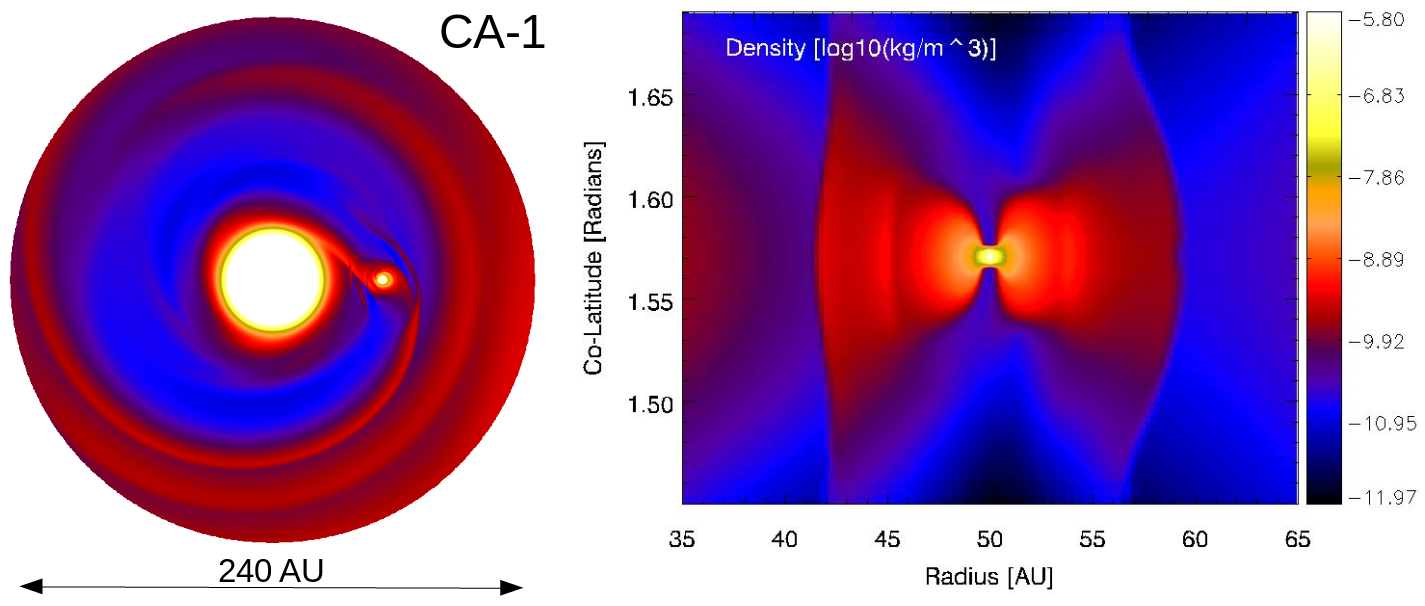}
\caption{The CA-1 simulation, where the 10 Jupiter-mass planet is formed via core accretion and is placed at 50 AU from the star. Left: the planet has opened a gap in the circumstellar disk. Right: zoom to the circumplanetary disk through a vertical slice.}
\label{fig:ca}
\end{figure*}

In the core accretion simulations, the CPD forms quickly while the 10 Jupiter-mass planet is built up through the mass-tapering function (see right-hand panel of  Figure \ref{fig:ca}). Because it is not possible to follow the entire core-accretion via hydrodynamic simulations, this initial fast planet augmentation is necessary in order to study the late stage of planet formation when a circumplanetary disk forms around the gas-giant. During this phase, the subdisk is still fed by a vertical gas influx from the circumstellar disk such as described in \citet{Szulagyi14}. The planet has opened a partial, eccentric gap in the gas of the protoplanetary disk (see left-hand panel of Fig. \ref{fig:ca}). 

\begin{figure}
\includegraphics[width=\columnwidth]{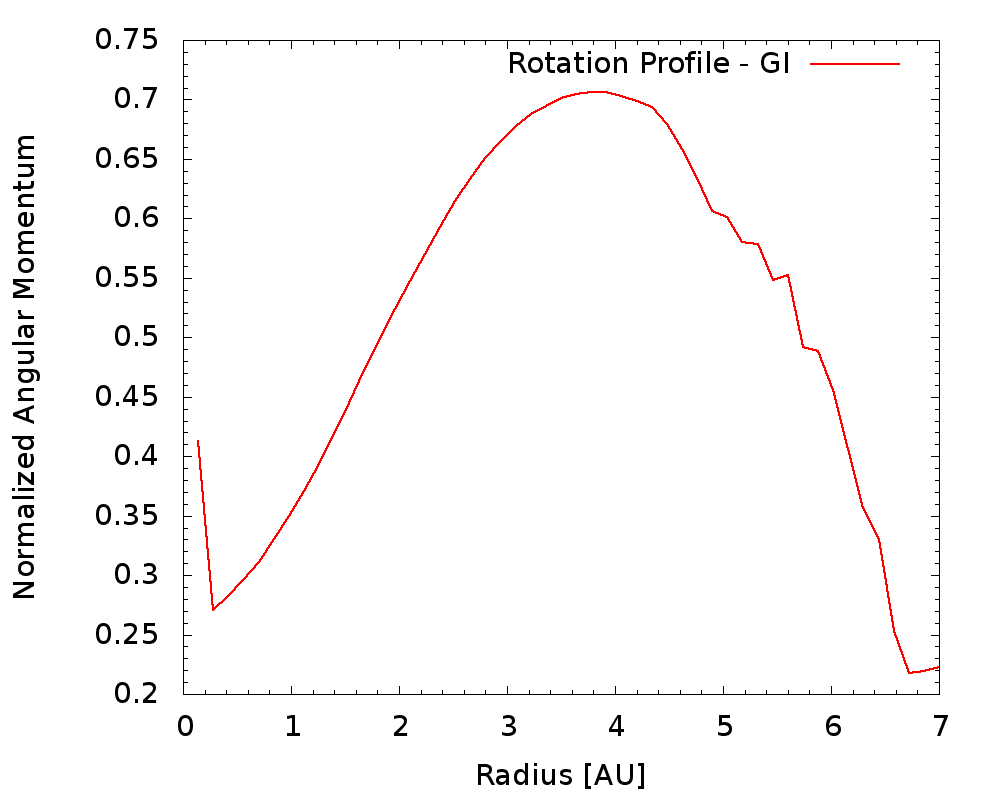}
\caption{Normalized angular momentum (i.e. rotational velocity divided by the local Keplerian velocity) profile on the midplane for the GI simulation. If the normalized angular momentum is one, that means perfectly Keplerian rotation, otherwise either sub-Keplerian ($<$1.0) or super-Keplerian rotation ($>$1.0). Clearly, rotation is dynamically important in the $\sim$ 2-6 AU  region; this is what we will define as circumplanetary disk in the following.}
\label{fig:gi_rotation}
\end{figure}

In the GI simulation the disk fragments into multiple clumps in the region at $60-80$ AU from the center after about 500 year, namely about
one disk rotation. While a detailed description of this and other similar simulations is deferred to a forthcoming paper (Mayer \& Quinn,
in preparation) here we focus on the formation of the circumplanetary disks. The clumps form with a wide range of masses, ranging between 2 and 20
Jupiter masses.  Some condense out of spiral arms in relative isolation while others appear to be triggered by a strong perturbation from
other clumps forming earlier (see also \citealt{AH99} and \citealt{Meru15}). The first 2-3 disk rotations after the onset of fragmentation
mark a highly chaotic phase in which protoplanetary clumps interact
vigorously among themselves and with the surrounding disk. The clumps lose mass via mutual 
tidal interactions and due to inward orbital migration, which in some cases appears to occur quickly, on the orbital timescale \citep{Malik15}, and simultaneously accrete mass from the disk. In all cases a CPD appears at the same time as the clump formation, as the subdisk results from
the higher angular momentum material accreted from the protoplanetary
disk that can reach centrifugal equilibrium around the denser core
that first collapses from the spiral arm \citep{Boley10}. The clear
dichotomy between an inner dense core and an outer much more diffuse
envelope, where rotation is dynamically important, can be seen on Figure \ref{fig:gi_rotation}. Here we show the normalized angular momentum (i.e. rotational velocity divided by the local Keplerian velocity) profile for the 10 Jupiter-mass clump soon after fragmentation. Clearly, the region between $\sim$ 2-6 AU has the largest rotation beyond the planet, this is what we will define as circumplanetary disk in the next Section. 

\begin{figure*}
\includegraphics[width=15cm]{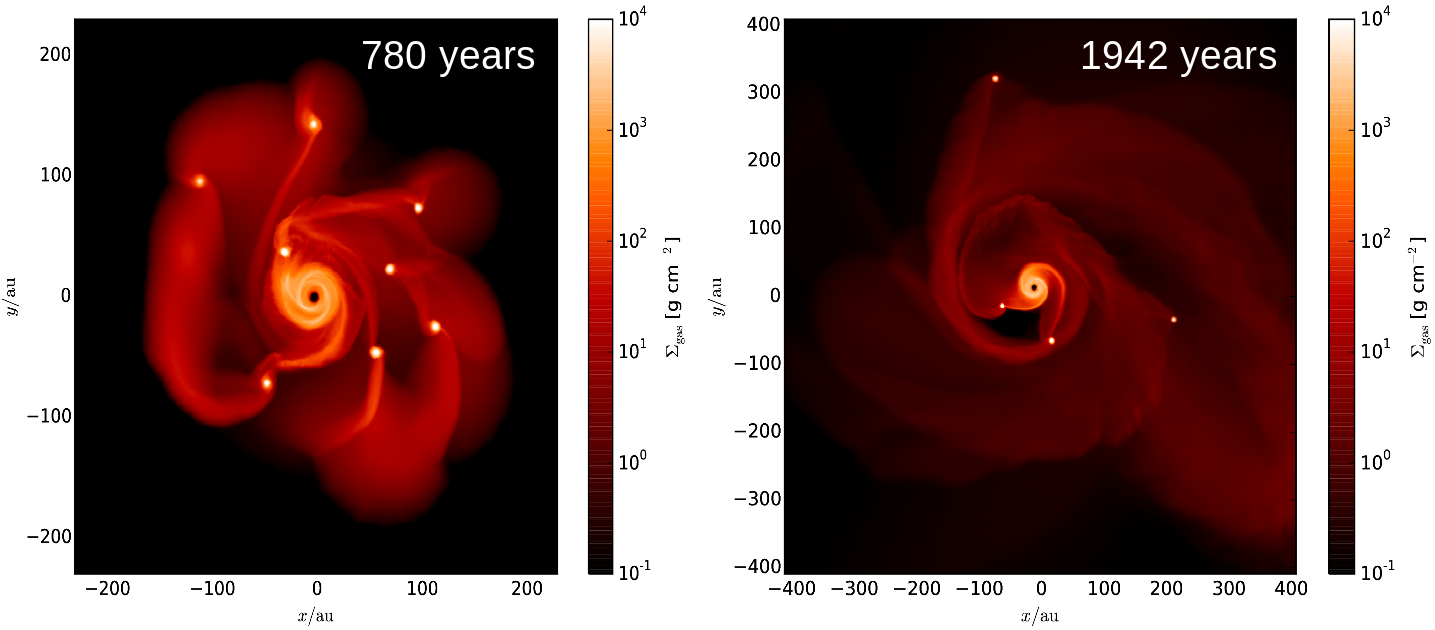}
\caption{Two snapshots of the GI simulation at 780 years and 1942 years. Only four clumps survive the first chaotic interaction phase between the clumps.}
\label{fig:gi_chaotic}
\end{figure*}

Figure \ref{fig:gi_chaotic} shows two snapshots of the GI simulation
in the early and late stage of the simulation, respectively. The
second snapshot shows only 4 clumps remaining among those initially
formed. Indeed, merger, inward migration and tidal mass loss are
responsible for disrupting about more than half of the initially
formed clumps. After $10^3$ years, the protoplanetary disk settles
into a more quiet phase as its Toomre $Q$ has risen enough to make it
relatively stable. At this stage we are left with a massive gas giant
of $\sim$ 10 Jupiter masses, an larger one on the order of 20 Jupiter
masses, and two even more massive objects that are clearly in the
brown dwarf regime. These clumps are on eccentric orbits and have
reached very high central densities at which dissociation would have
already begun if included (see Section \ref{sec:discussion}). Indeed,
the densities are comparable to those in \citet{Galvagni12} before the
onset of dissociation, which is not included here and would not be
reached anyway since we shut-off the cooling in the core well before
it reaches that density. Following dissociation, the core would
collapse dynamically in timescales of years to the density of Jupiter \citep{Bod89,Helled06}. This ``second collapse'' phase would occur below our resolution limit,  hence it cannot be followed here. It is therefore likely that these protoplanets and proto-brown dwarfs will survive indefinitely even if they migrate to less than an AU
from the star, although most of the CPD could be stripped in that case
(see Discussion Section).

The subdisk around the GI formed 10 $\mathrm{M_{Jup}}$ protoplanet has grown in mass during the interactions of the clumps, but the ratio between CPD mass 
and protoplanet mass has remained roughly constant, only increasing slightly (see Sect. \ref{sec:timeevol}). We will focus our analysis, in the rest of the paper, on the 
lowest mass object, the gas giant with mass around 10 Jupiter-masses;
however we also have studied the other clumps to confirm that the results presented here on the properties of the CPD are general. 

\subsection{Defining the CPD boundaries}
\label{sec:cpd_def}

To obtain the mass of the circumplanetary disk, we first need to
define its boundaries. There are three main ways to define the boundary, namely:
\begin{enumerate}
\item draw streamlines and account for the area where the flow is bound to the planet (i.e. orbiting around it)
\item compute the eccentricity of the orbit of a fluid-element at
various radii from the planet, then use the circular orbits to define the boundaries; however, in case of massive planets --- such as in this work --- the CPD can become eccentric, so this method would not be suitable
\item calculating the normalized angular momentum around the planet,
meaning the z-component of the angular momentum normalized by the local Keplerian velocity at a given radius; then setting a minimum value  --- i.e. how sub-Keplerian the gas is in the CPD --- sets the boundaries. 
\end{enumerate}

The work of \citet{Szulagyi14} showed that the first and third methods
lead to roughly the same, 0.5 $\mathrm{R_{Hill}}$, CPD radius in the
case of a 1 $\mathrm{M_J}$ planet at 5.2 AU. In this work, however, we
define the CPD boundaries via the normalized angular momentum, because
comparing the GI and CA simulations with this quantity is particularly
useful. We decided on a 45\% minimum Keplerian rotation to define the
boundaries of the subdisks.  Therefore the mass integral within this
region, ---where the normalized angular momentum is larger than 0.45
--- in all the different simulations can lead to a valid comparison of
the CPD-masses.  Furthermore, we checked the streamline method, and we
get roughly the same radius for the CPD as that from the $>0.45$
normalized angular momentum value.

Because the definition of the CPD borders is still arbitrary, we also
compared the mass of the entire Hill-spheres.  The CPD is definitely a subset of the Hill-sphere, and the Hill-sphere is easily definable with $\mathrm{R_{Hill}}=a_p(M_p/M_*)^{1/3}$; therefore, the comparison of the Hill-sphere masses can eliminate any possible uncertainty of the CPD mass comparisons due to the arbitrary subdisk borders.

\subsection{Comparing the Density profiles and Masses}
\label{sec:dens}

\begin{figure}
\includegraphics[width=\columnwidth]{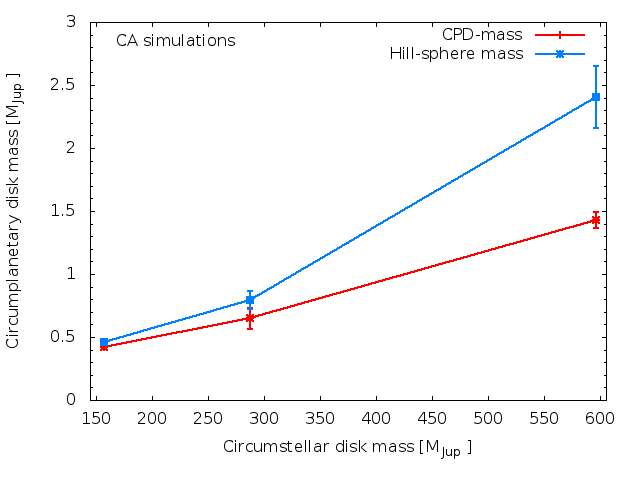}
\caption{The circumplanetary disk masses from the core accretion
simulations as a function of the circumstellar disk masses. The CPD
mass seems to scale linearly with the protoplanetary disk mass, hence
even core accretion planets can have a very massive subdisk mass (12\%
of the planetary mass in the case of the 600  $\mathrm{M_{Jup}}$
protoplanetary disk). For comparison, we also show the mass in the
Hill sphere in each case; again, the relationship with the circumstellar disk mass is roughly linear.}
\label{fig:cpd_vs_csd}
\end{figure}

As mentioned in the previous section, when calculating the masses of
the CPD, we integrated the mass where the rotation of the gas is at
least 45\% Keplerian. From the CA simulations --- since all planets
were 10 $\mathrm{M_{Jup}}$ --- we compared the CPD masses with the
circumstellar disk masses (see
Fig. \ref{fig:cpd_vs_csd}). The error bars were calculated as the standard deviation of 10 outputs of the simulation over one orbit of the planet. Surprisingly, even in the case of
radiative simulations the CPD mass seems to (nearly) scale linearly
with the circumstellar disk mass, with the relation
\begin{eqnarray}  
\mathrm{M_{CPD}}=\mathrm{M_{CSD}}\cdot (2.26\pm0.12)\cdot 10^{-3}+(6.49\pm 2.37)\cdot 10^{-2}
\end{eqnarray}
This linear relationship is very important especially for observations
aiming to detect the CPD, because it means that the mass of the
subdisk is not necessarily related to the mass of the planet, rather,
more massive circumstellar disks will have more massive
circumplanetary disks. Therefore, observations should not target very
massive gas-giants to detect the subdisk, but instead target massive
circumstellar disks where the planet has opened a gap (and therefore
the gap region is optically thin).

We also compared the entire Hill-sphere masses with the circumstellar disk mass (Fig. \ref{fig:cpd_vs_csd}). The relationship is again linear:
\begin{eqnarray}
\mathrm{M_{Hill}}=\mathrm{M_{CSD}}\cdot (4.57 \pm0.60)\cdot 10^{-3} -0.37(\pm 0.24)
\end{eqnarray}
The Hill sphere to CPD mass ratios scale from 1.1 to 1.7 for the CA
simulations, and, more massive circumstellar disks have larger mass ratios.

Because the CPD masses scale with the circumstellar disk mass, in our
0.6 $\mathrm{M_{solar}}$ circumstellar disk the subdisk was 1.2
$\mathrm{M_{Jup}}$, giving a CPD-to-planet mass ratio of 12\%. This is
a significantly higher ratio than found so far in CA simulations,
$10^{-4}-10^{-3}\mathrm{M_{planet}}$ \citep{Gressel13,AB09b,Szulagyi14,Szulagyi16a}.
Now it is understandable that the reason for the discrepancy is that
those works all used very light circumstellar disks
($\sim10 \mathrm{M_{Jup}}$), so the CPD is correspondingly less
massive.  So far, the GI subdisk simulations predict 25\%
$\mathrm{M_{planet}}$ \citep{SB13} and 50\%
$\mathrm{M_{planet}}$  \citep{Galvagni12}. Comparing with these
values from GI simulations, the CPD-to-planet mass ratios in CA
simulations are lower, not by several orders of magnitude, but only by
a factor of eight. Therefore, it cannot be said that GI formed CPDs are definitely more massive; it will depend on the circumstellar disk mass. Thus, observationally, the CA and GI formation mechanisms cannot be distinguished with confidence solely from the observed CPD masses. 

The subdisk masses in our GI simulations reach values even higher than
the aforementioned 25-50\% of the planetary mass.  Applying our
normalized angular momentum threshold of 0.45 we find that soon after
the clump forms the CPD mass is about 6 $\rm{M_{Jupiter}}$ compared to
10 $\rm{M_{Jupiter}}$ for the protoplanetary core inside it. At the
latest time (corresponding to the snapshot on the right of
Figure \ref{fig:gi_chaotic}), the subdisk grows to about 10
Jupiter-masses while the protoplanetary core, which has continued to
collapse, has grown only to about 13 $\rm{M_{Jupiter}}$. Hence, the
CPD-to-planet ratio is roughly 60\% of the protoplanet mass at the
beginning, in substantial agreement with the results
of \citet{Galvagni12}, while at later times it becomes comparable to
the protoplanet mass. We note that at late times the clump has
acquired a very eccentric orbit, moving out to $R > 150$ AU and
gathering high angular momentum gas from the outer fringes of the
disk.  Accretion along this outgoing  orbit might explain the increasing CPD mass with time relative to previous work. (We note, in particular, 
that the collapsing clumps studied in \citealt{Galvagni12} were
isolated hence the interplay between accretion and orbital evolution
was missing by construction.)

\begin{figure}
\includegraphics[width=\columnwidth]{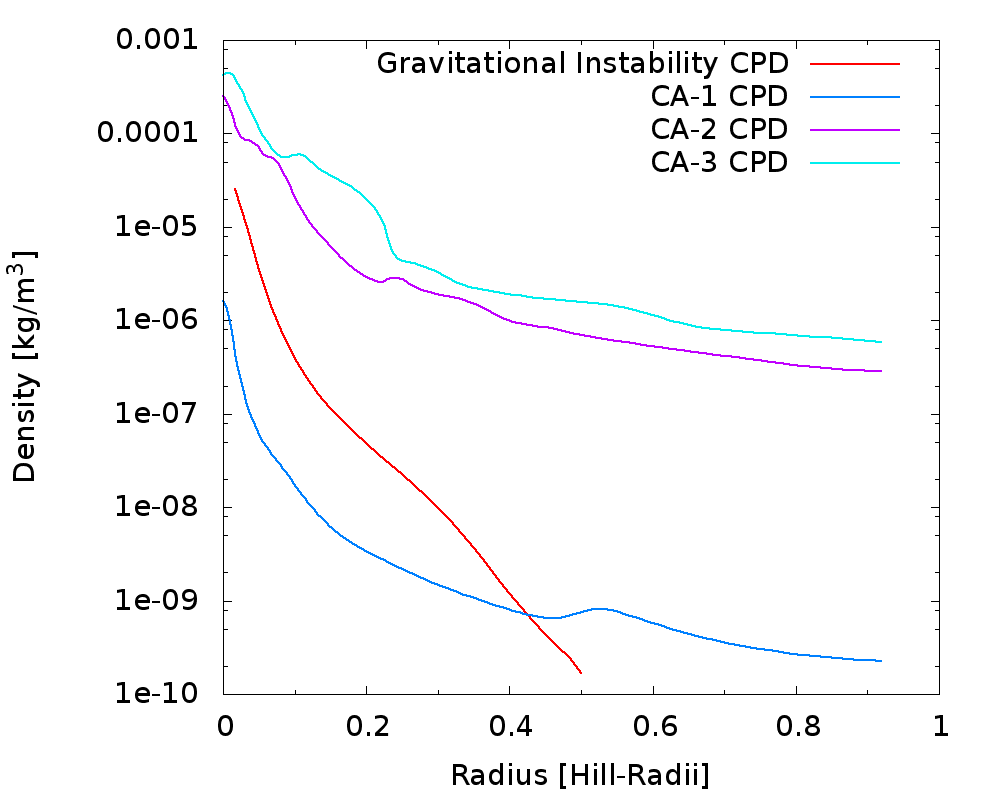}
\caption{The midplane density profiles of the different simulations
from the planet (left-hand side) to 1 Hill-radius. The CA-2 and CA-3
simulations have higher, and the CA-1 calculation has lower volume density in the Hill-sphere than the gravitational instability simulation.}
\label{fig:densprof}
\end{figure}

We also compared the midplane density profiles of the CA and the GI
simulations, see Fig. \ref{fig:densprof}. The CA-2 and CA-3
calculations predict larger volume densities in the midplane than the
gravitational instability calculations, while the CA-1 simulations
gives the lowest density in the midplane. This is due to the fact that
the core accretion simulations are more compact: the planet is a
point-mass, and the circumplanetary disk is not as extended as in the
GI case. In the GI simulations, the planet is not a point mass, but an
extended clump which is still collapsing.
Fig. \ref{fig:densprof} is plotted with respect the Hill-radius, but
the Hill-spheres are significantly different in physical size in the
various simulations. This is the reason why the GI simulation has
lower or comparable volume density in the midplane, but an overall
more massive Hill-sphere, than in the CA simulations. 

As we discuss later in the Discussion section, with increased resolution, and with the inclusion of molecular dissociation in the GI simulations, the core is expected to collapse into a fully fledged planet of a few Jupiter radii in $< 10^5$ years \citep{Helled14}, as hinted
by the isolated collapse simulations of \citet{Galvagni12}. However,
what fraction of the mass would actually collapse to this final state depends on the angular momentum profile at small radii. In \citet{Galvagni12} the angular momentum transport from the core to the CPD was occurring due to non-axisymmetric instabilities, which appear not to be captured yet in our global simulations as we limit the cooling above a certain density. 
Resolving angular momentum transport processes is important in order
to answer the following question; when exactly will the
protoplanetary core become compact enough to be similar to the planet
configuration in the CA simulations? When this happens, one would
expect the CPD to evolve towards a state similar to the subdisk in the
CA simulations. However, there is one aspect that will prevent the two
scenarios from converging, namely the fact that the clump in the GI simulations has a significantly
higher angular momentum budget. In the late stage, the total angular
momentum of the subdisk in the GI simulation is about an order of
magnitude higher than in the CA-1 simulation. Nevertheless, the
specific angular momentum is comparable in the two cases, indicating
that they are both built from material accreted from the outer
circumstellar disk. The much larger CPD mass in the GI simulation ($11
M_{Jup}$ as opposed to $0.5 M_{Jup}$ in the CA-1 computation) creates
a major division for the subsequent dynamical evolution. In the GI case, the protoplanetary core and the CPD will continue to collapse together, while in the CA, the subdisk will accrete onto an already compact planet. 

\subsection{Comparing the Temperature Profiles}

\begin{figure}
\includegraphics[width=\columnwidth]{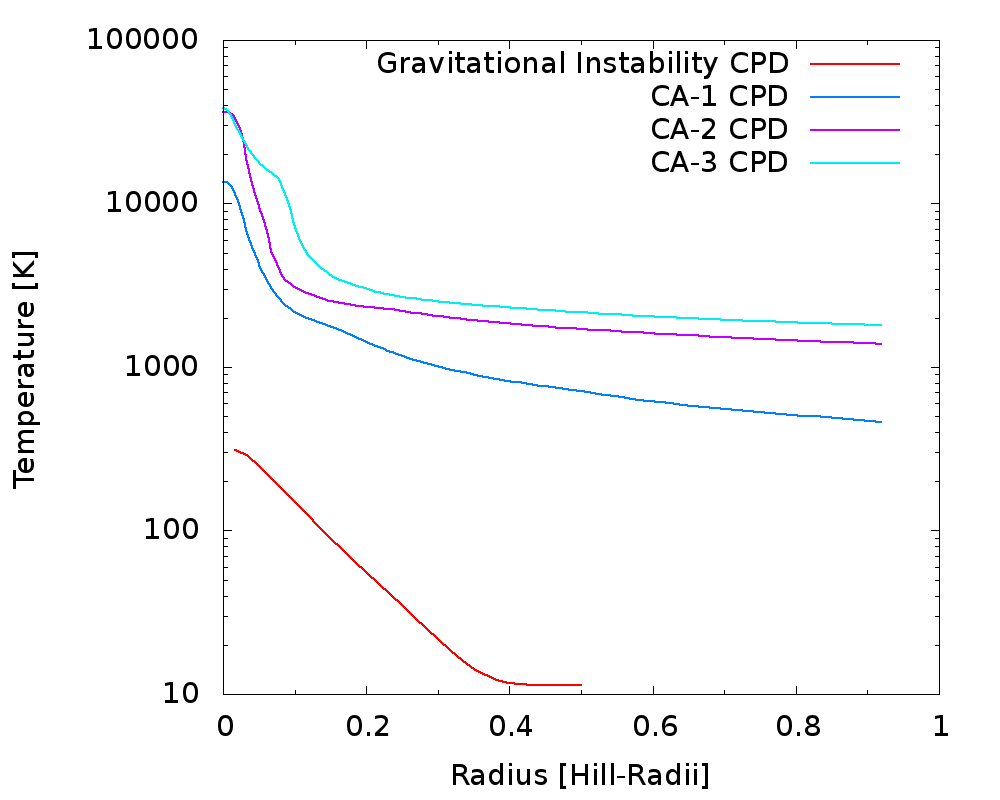}
\caption{The midplane temperature profiles of the different simulations from the planet (left-hand side) to 1 Hill-radius. Clearly, all the core-accretion simulations predict more than an order of magnitude higher temperature than the gravitational instability simulation.}
\label{fig:tempprof}
\end{figure}

We compared the midplane temperature profiles of the three core-accretion simulations and the gravitational instability calculations inside the Hill-sphere (Fig. \ref{fig:tempprof}). We found more than an order of magnitude difference in the bulk temperature between the core-accretion and the gravitational instability predictions. The latter predicts a characteristic temperature of $\sim 50$ K in the circumplanetary disk (between $\sim$ 0.1 and 0.3 Hill-radii), while the core-accretion simulation with the planet at 50 AU predicts 800-1000K inside the circumplanetary disk defined in Sect. \ref{sec:cpd_def}. 

If we compare the various core-accretion simulations with each other,
we see that the Hill-sphere gas has a higher bulk temperature when the
planet is at 5.2 AU in contrast with the 50 AU simulation. However,
the difference does not come from the different semi-major axes,
partially because we did not use stellar irradiation in these
calculations. Instead, the difference in temperature is due to
different circumstellar disk masses. In all the calculations we used a
dust-to-gas ratio of 0.001; therefore, the amount of the integrated
dust in the disk is also higher when the  circumstellar disk mass is
higher. Dust is the main heating source in protoplanetary disks,
because the more dust we have, the greater the optical depth of the disk;
hence, the cooling is less efficient. Even though our temperature of the CPD coincides with the drop in opacity at the dust sublimation temperature assumed in the \citet{BL94} opacity table, the CPD is optically thick even across this opacity drop due to dust sublimation and the high density. If we compare the temperature profiles of this work and \citet{Szulagyi16a} where the Jupiter-mass planet is embedded in a $\sim$10 $\mathrm{M_{Jupiter}}$ circumstellar disk, we see that the temperatures are significantly lower there. 

When comparing the temperatures of two different simulations,
especially with two different methodologies (here grid based and
smoothed particle hydrodynamic computations), it is important to
understand how the temperature is affected by the numerics. As we
described in Sect. \ref{sec:numerical}, the core-accretion simulations
were carried out with the flux limited diffusion approximation, while
the GI calculations were done with a phenomenological cooling law that
was calibrated to flux limited diffusion results with the same
code. Another important factor for the temperature calculations is the
resolution. Our hydrodynamic resolutions are comparable, the GI
resolution being $10^{-2}$ AU while the core-accretion is
$3\times10^{-2}$ AU for the planet at 50AU (simulation CA-1) and
$3\times10^{-3}$ AU for the planets at 5.2 AU. The gravitational
softening in the GI case was 0.01 AU, while it was 0.17 AU for the
CA-1 simulation and 0.017 AU for CA-2 and CA-3. Therefore, the
comparable resolutions and gravitational softenings provide valid
comparisons for the temperature.

\begin{figure}
\includegraphics[width=\columnwidth]{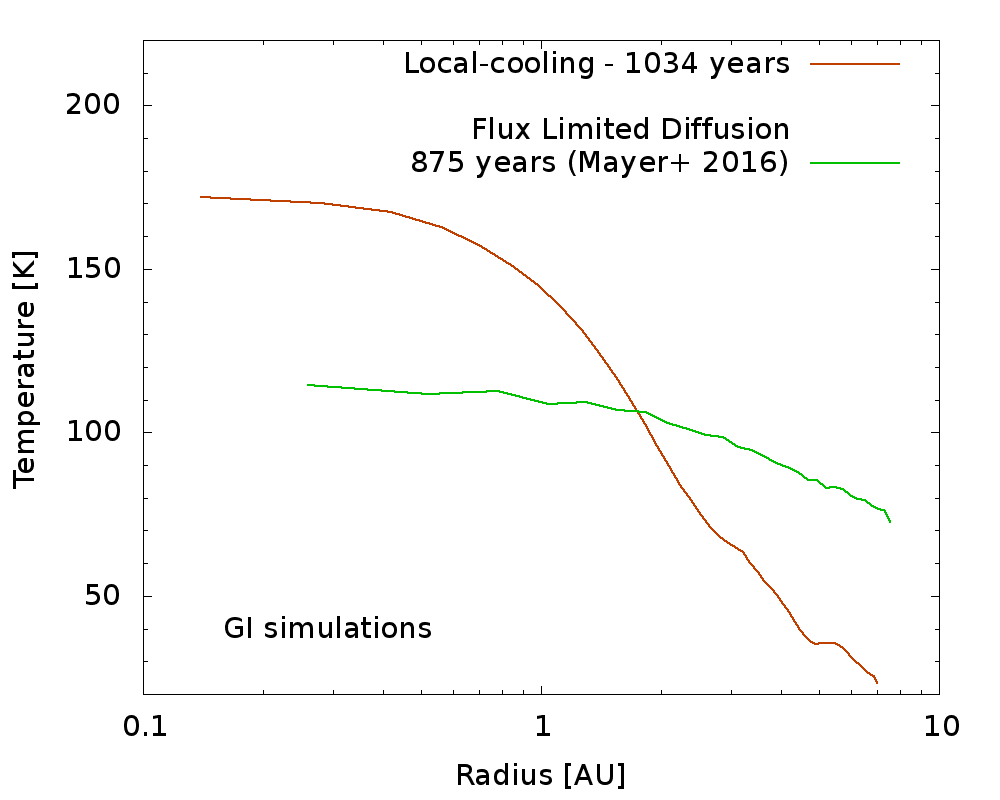}
\caption{Comparison of the clump temperature profiles with a flux
limited diffusion simulation from \citep{Mayer16} (green) and this
work (red) with local cooling. Clearly, the full radiative transfer
with flux limited diffusion also gives very low clump and subdisk
temperatures, similar to what was found in this work. Hence, the more
than an order of magnitude temperature difference found between our GI and CA simulations is robust.}
\label{fig:RT_beta}
\end{figure}

In order to check whether the lack of the flux limited diffusion
approximation in the GI simulation has an effect on the temperature,
we did a comparison with a similar simulation with the flux limited
diffusion approximation included \citep{Mayer16} that was carried out
with a similar SPH code (see Figure \ref{fig:RT_beta}). The clump
from \citet{Mayer16} had a mass of 8 $\mathrm{M_{Jupiter}}$, so
similar to our 10 $\mathrm{M_{Jupiter}}$ protoplanet. Due to the
inclusion of the flux limited diffusion approximation
in \citet{Mayer16}, the resolution is lower, and the simulation
timespan is shorter than in this work. Nevertheless, as it can be seen, the comparison result is re-assuring as the temperature in the outer region corresponding to the CPD is below $100$ K, significantly lower than the flux limited diffusion core accretion simulations.

The reason for the temperature difference between the CA and GI
simulations is twofold. First, the optical depth is, of course,
playing a large role in determining the cooling rate. Because the CA-2
and CA-3 simulations have larger densities close to the planet than
the GI calculation (see Sect. \ref{sec:dens}), the gas is more
optically thick, and it cools less efficiently than in the GI case. 
In the GI case, optical depths are of order unity in the CPD region
(but increasing to $> 1000$ in the core), since they reflect the
conditions necessary in the disk for gas to be able to fragment and
form a  clump, i.e. the cooling time has to be a few times the local
orbital time \citep{Gammie01,Rafikov03, CL09}. Secondly, the profile
of the gravitational potential well --- i.e. the size of the
protoplanet --- is also significantly affecting the temperature. In
the GI simulations the protoplanet has initially a size of few AU
before it begins to collapse. On the other hand, in the CA
calculations, an entire 10 Jupiter-mass planet is compressed into a
point mass with a gravitational softening of 0.17 AU or 0.017 AU, for planets at 50 AU and 5.2 AU respectively. This means that the gravitational potential well is narrower and deeper in this case than in the GI simulations. Therefore, the gas can release more energy into heat while accreting to the planet. This can be understood with the accretional luminosity formula:
\begin{equation}
L_{acc}=\frac{GM_{p}\dot{M_{p}}}{R_{p}}
\end{equation}
where a smaller planetary radius, $R_p$, gives a larger accretional
luminosity for the same planet mass, $M_p$, and the same accretion
rate, $\dot{M_{p}}$. 
As we discussed in the previous section, in the GI simulation,
eventually the protoplanetary core would collapse to a few Jupiter
radii, which suggests the system will begin to look more like the CA
case in terms of gravitational potential profile, and thus, the
release of heat as gas flows inward would also become more
similar. This would imply that the CPD temperatures should become
alike in the GI and CA cases. However, we pointed out in the previous
section that the angular momentum budget of the planet plus CPD  is significantly different in the two cases. This implies that the collapse will be different at all times. In the GI simulation, the CPD will shrink more slowly with time as angular momentum has to be removed at all radii in order for the collapse
to proceed, and it may even gain angular momentum from the collapsing core due to angular momentum transport via bar-like and spiral instabilities. \citet{Galvagni12} found that when the latter happens, the CPD becomes nearly Keplerian and reaches centrifugal equilibrium around the protoplanetary core. As a result the CPD temperature remains low, below 100 K. While this late-stage evolution will have to be re-addressed with new, even higher resolution global simulations, it already suggests that the remarkable difference in CPD temperatures between the two formation mechanisms should persist on long timescales, well beyond $10^4$ years.

\subsection{Time-evolution of the disk instability simulations}
\label{sec:timeevol}

In the case of the gravitational instability simulations, a steady
state cannot be reached by the end of the simulations. A clump of a
few Jupiter masses is expected to collapse into bona-fide gas giant of
about a Jupiter radius in $10^4 - 10^5$ years, depending on the
angular momentum, the metal enrichment during the collapse, the mass
accretion rate from the disk, and other
conditions \citep{Helled14}. The collapse timescale is generally
defined as the time it takes to reach $\mathrm{H_{2}}$ dissociation,
which triggers a dynamical collapse. It is numerically very
challenging to follow the collapse all the way with hydrodynamic
simulations, partially because the more compact the clump, the slower
the computation. We managed to follow the GI collapse for almost a hundred CPD dynamical times, thanks to access of one of the fastest supercomputer in the world. Once the inner dense core of the clump contracts to a couple of  gravitational softening lengths ($\sim 0.02$ AU) the collapse is artificially halted in our GI simulation, but this is not an issue for studying the subdisk.

\begin{figure}
\includegraphics[width=\columnwidth]{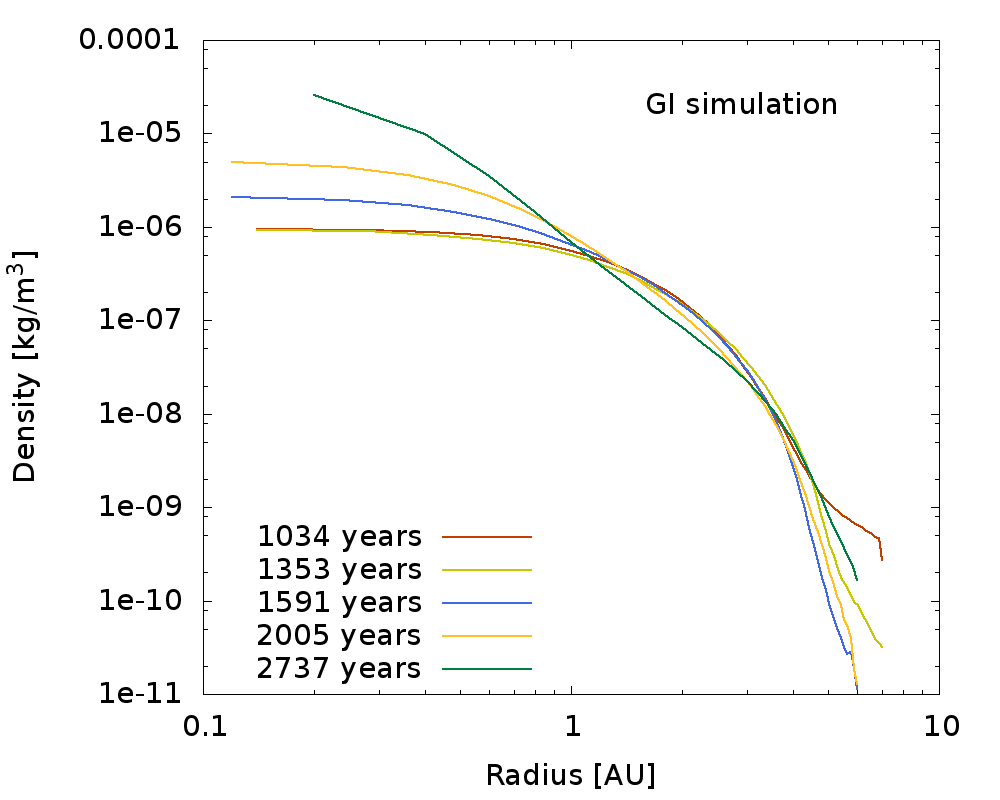}
\caption{The time evolution of the density on the midplane in the GI simulation. }
\label{fig:densevol}
\end{figure}

Figure \ref{fig:densevol} shows the time evolution of the clump's
density in the midplane between 1034 and 2736 years. As the clump
collapse to form the  planet, the peak density in the center
increases, while in the outer parts of the clump the density
decreases.  Understanding the time evolution of the simulation is
important when comparing it with the CA simulations, where steady
state has already been reached\footnote{Note that accretion is still
ongoing in the CA simulations; therefore, the density does increase in the innermost cells around the planet point-mass}. 

\begin{figure}
\includegraphics[width=\columnwidth]{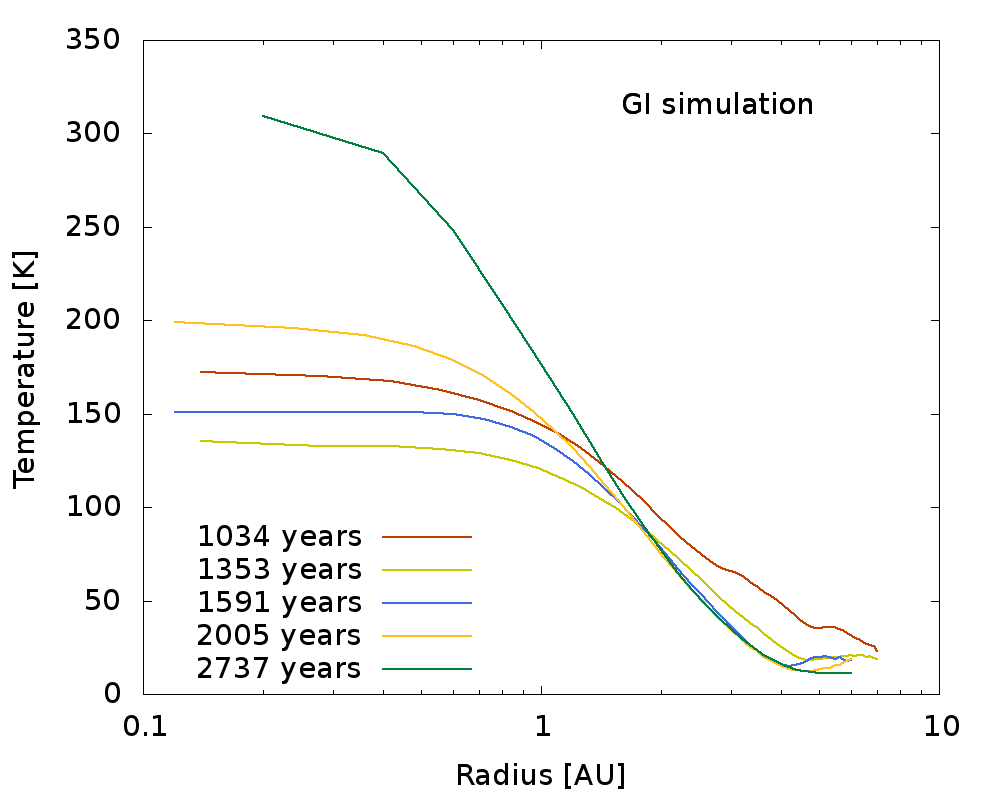}
\caption{The midplane temperature profiles of the different simulations from the planet (left-hand side) to 1 Hill-radius. Clearly, all the core-accretion simulations predict more than an order of magnitude higher temperatures than the gravitational instability simulations.}
\label{fig:tempevol}
\end{figure}

We also show on Fig. \ref{fig:tempevol} how the temperature changes
during the collapse of the clump. While the temperature rises in the
central parts (i.e. the interior of the protoplanet) by $\sim$ 160 K
over 1700 years, the outer parts of the clump, what we call the
circumplanetary disk remain roughly at the same temperature
($\sim20-60$K). This gives us a robust comparison of temperature with
the core-accretion simulations, because the temperature remains the
same in the GI circumplanetary disk, irrespective of the
collapse. Therefore, the fact that we cannot follow the birth of the
planet --- i.e. the full collapse of the clump --- does not change the
conclusion regarding the orders of magnitude difference in circumplanetary gas temperature between the CA and GI simulations.

\section{Discussion}
\label{sec:discussion}

The time-scales of core-accretion and disk instability formation
mechanisms are very different: the former takes place on $10^5-10^6$
years, while the latter is $\sim10^4-10^5$ years long. Therefore, our
two sets of simulations describe a slightly different epoch of the
circumstellar disk. Our GI simulation represents a very early stage in
the circumstellar disk evolution, perhaps during the Class I phase, while the core-accretion simulations describing an epoch when the circumstellar disk is slightly more evolved, late Class-I to mid-Class II phase.

Our results, of course, have uncertainties because of the differences
between the grid-based code and the smoothed particle hydrodynamic
code. In the former, the ionization and dissociation is not included, because its numerical implementation is stable only with specific, low-order Riemann-solvers which are not the method of choice in our code \citep{Vaidya15}; therefore, it is possible that temperatures are overestimated. 
On the other hand the GI simulation does not include flux limited
diffusion approximation radiative transfer. Nevertheless, we tested
the temperature of the GI simulation by comparing it to a low
resolution flux limited diffusion approximation simulation and
confirmed that they result in a very minimal difference for the CPD temperature. In this respect, our conclusion concerning the major difference between the early stage of the CPD in the GI and CA
simulation is robust.

One may wonder how our conclusion would change during the subsequent
evolution in the GI case after our simulation is ended, since the
collapse of the planet is continuing and steady state was not
reached. However, there is another key difference between our two sets
of simulations that keeps our conclusion valid on a longer timescale:
the total angular momentum budget of the planet plus subdisk
system. Indeed, as we have have explained in Section \ref{sec:dens},
the CPD in the GI case has more than an order of magnitude higher
angular momentum than the corresponding CA case. We also checked that
the GI CPD, due to its large radius and low density, is Toomre stable,
with $Q > 1.5$ at all times. This implies that rapid dynamical angular
momentum transport via self-gravitating instabilities will not
operate, but a more gentle transport can happen by global
non-axisymmetric instabilities (see in \citealt{Galvagni12}). The
$\sim 10 M_J$ core in the GI simulation would contract to a few a
Jupiter radii in $< 10^5$ years, during which time the CPD can reach
centrifugal equilibrium due to internal angular momentum transport
between the collapsing core and the subdisk. If the CPD is threaded by
a magnetic field, it may become magneto-rotationally unstable \citep{Turner14,Fujii14}, but for
realistic values of viscosity ($\alpha < 0.01$), the evolutionary
timescales will be long ($> 10^5$ years). Hence we can safely conclude
that the CPD by itself is a long-lasting structure despite the fact
that the inner protoplanetary core has not reached steady state and will continue to collapse further. Instead, what will likely happen
is that part of the CPD could be stripped by tides if the clump
plunges inward on its eccentric orbit and reaches less than 10
AU. Therefore, we predict that for GI clumps, either an extended, cold
CPD is present for up to $10^5$ years, or there is no CPD if it has
been lost by tides during migration.  In no case do we expect a hot CPD akin to that in the core-accretion case.

In the case of our Solar System, where the giant-planets were most
likely formed via core-accretion, the integrated mass of Jupiter's and
Saturn's satellites makes up $2\times 10^{-4}$ of the planetary
mass. Assuming the interstellar medium value for the dust-to-gas ratio
(i.e. 1\% dust), this would mean a minimum mass for the CPD of 2\% of
the masses of our two largest gas-giants. However as \citet{CW02}
pointed out, this mass has to be processed during the entire satellite
growth timescale, i.e. it does not all have to be present at one given
instant of time. The reason is that the  CPD is not a closed reservoir
of mass, unlike the  circumstellar disk. The subdisk is fed by the
circumstellar disk, and loses mass through the accretion onto the
planet. Therefore, the subdisk mass is not constant in time, but
depends on the feeding and mass loss balance, and it changes as the
circumstellar disk evolves and the planet grows. Given that our Sun
probably had a rather low mass circumstellar disk, the circumplanetary disks around our gas-giants must have had low masses too. In other planetary systems, however, where the protoplanetary disk mass is higher (at present, or at earlier stage), the CPD mass can be also higher and can result in more massive, more extended satellite systems.

In the disk instability model, there might also be a second generation
of subdisks forming around the fully fledged giant planets. Indeed, if
protoplanets migrate inward to $< 10$ AU, their Hill radius will
shrink; therefore, the CPDs can be stripped away by tides almost
entirely \citep{QT98}. In this case, only a dense protoplanetary core
is left, which can  survive even at orbital radii of $\sim$1 AU at the
densities found at the end of our GI simulation. Since massive
protoplanets in massive self-gravitating disks can migrate inward on
timescales of $< 10^5$ year, we argue that detection of the original
CPDs formed by disk instability is more likely in the early
evolutionary phase of the protoplanetary disk, before the Class II
stage. Later, a new subdisk might be accreted by the newly formed gas
giant, but it will be much more compact than the first population of
GI subdisks. These second generation of CPDs probably will have
thermodynamic properties analogous to core-accretion subdisks, given
that they formed around fully fledged giant planets. If such a second generation exists, then the CPDs between the two formation mechanism will not likely differ much.

\section{Conclusion}   

In this paper we compared the main characteristics (mass \& temperature) of circumplanetary disks around core accretion and gravitational instability formed gas giants. We used state-of-the-art hydrodynamic simulations with as similar initial parameters as possible to reveal the key differences between the subdisks of the two main planet formation scenarios. 

The core accretion simulations were carried out with the JUPITER code,
featuring a radiative module with the flux limited diffusion
approximation and mesh refinement. The disk instability simulations
were performed with the ChaNGa smoothed particle hydrodynamic code,
matching the resolution of the grid-based simulations and having a
radiative cooling calibrated to flux limited diffusion results.  We ran three core accretion and one disk instability simulation with 10 Jupiter-mass planets in massive circumstellar disks (158, 290 and 600 $\mathrm{M_{Jup}}$). In two core accretion simulations the planets had a semi-major axis of 5.2 AU, the third simulation featured a gas-giant at 50 AU distance from its star. In the GI calculations the semi-major axis was also 50 AU for our chosen, 10 Jupiter-mass protoplanet, although the orbit varied a bit through interactions with other clumps.

We found from the core-accretion simulations that the subdisk mass
linearly scales with the circumstellar disk mass, even in these
radiative simulations. This means that core accretion CPDs can be
nearly as massive as their GI counterparts, if the protoplanetary disk
has the same mass. In the 0.6 $\mathrm{M_{solar}}$ circumstellar
disks, the CA simulation resulted in a CPD with a mass of 12\%
$\mathrm{M_{p}}$, while we found a CPD mass of 50\%-100\%
$\mathrm{M_{p}}$ in the GI computation. Previous works predicted a 4-5
orders of magnitude mass discrepancy, but we were able to show that
was because of their orders of magnitude differences in circumstellar disk masses.

On the other hand, our finding is that the temperature differs by more than an order of magnitude between the GI and CA formed CPDs. According to the simulations, the bulk subdisk temperature is $<$ 100 K in the case of disk instability, and over 800 K for all the CA computations presented in this paper. The reason for this discrepancy lies in the different gravitational potential wells and opacities. Because the protoplanet is a few AU wide extended clump in the GI simulations, while it is a fully formed giant planet with a radius of 0.17 AU (meaning the gravitational potential smoothing length) in the CA-1 simulation, the accreted gas has significantly more energy to release into heat in the latter case than in the former. 

The large temperature contrast between CA and GI circumplanetary disks provides a convenient tool for observations on young, embedded planets to distinguish between the two main formation mechanisms.

% -------------------------------

\section*{Acknowledgments}

We thank for the anonymous referee for the thoughtful review, which helped to improve the paper. J. Szul\'agyi acknowledges the support from the ETH Post-doctoral Fellowship from the Swiss Federal Institute of Technology (ETH
Z\"urich). T. Quinn was supported  by NASA grant NNX15AE18G. This work has been in part carried out within the frame of the National Centre for Competence in Research  ``PlanetS"  supported by  the  Swiss  National Science Foundation. Computations have been done on the ``M\"onch" and ``Piz Dora" machines hosted at the Swiss National Computational Centre.

\label{lastpage}
%http://iopscience.iop.org/0004-637X/526/2/1001/pdf/40637.web.pdf

\end{document}